*Cultural Dimensions of Artificial Intelligence Adoption: Empirical Insights for Wave 1 from a Multinational Longitudinal Pilot Study*

Michelle J. Cummings-Koether 1*; Franziska Durner 2*; Theophile Shyiramunda 3*; Matthias Huemmer 4*

**1* Prof. Dr. Michelle J. Cummings-Koether,**
Professor, Institute for the Transformation of Society (I-ETOS), Deggendorf Institute of Technology - European Campus Rottal-Inn, Germany
E-Mail: michelle.cummings-koether@th-deg.de
ORCID ID: https://orcid.org/0000-0002-7137-3539

**2* Franziska Durner, MA,**
Research Associate, Institute for the Transformation of Society (I-ETOS), Deggendorf Institute of Technology - European Campus Rottal-Inn, Germany
E-Mail: franziska.durner@th-deg.de
ORCID ID: https://orcid.org/0000-0002-8064-5364

**3* Dr. Theophile Shyiramunda,**
Postdoctoral Research Associate, Institute for the Transformation of Society (I-ETOS), Deggendorf Institute of Technology- European Campus Rottal-Inn, Germany
E-Mail: theophile.shyiramunda@th-deg.de
ORCID ID: https://orcid.org/0000-0001-6725-3756

**4* Prof. Dr.-Ing. Matthias Hümmer,**
Professor, Institute for the Transformation of Society (I-ETOS), Deggendorf Institute of Technology - European Campus Rottal-Inn, Germany
E-Mail: matthias.huemmer@th-deg.de
ORCID ID: https://orcid.org/0009-0003-8122-470X

**Corresponding author @ Dr. Theophile Shyiramunda,**
Postdoctoral Research Associate, Institute for the Transformation of Society (I-ETOS), Deggendorf Institute of Technology- European Campus Rottal-Inn, Germany
E-Mail: theophile.shyiramunda@th-deg.de
ORCID ID: https://orcid.org/0000-0001-6725-3756

## Declarations

**Conflict of Interest**
The authors have no relevant financial or non-financial interests to disclose.

**Funding**
No funding has been received

# Abstract

The swift diffusion of artificial intelligence (AI) raises critical questions about how cultural contexts shape adoption patterns and their consequences for human daily life. This study investigates the cultural dimensions of AI adoption and their influence on cognitive strategies across nine national contexts in Europe, Africa, Asia, and South America. Drawing on survey data from a diverse pilot sample (n = 21) and guided by cross-cultural psychology, digital ethics, and sociotechnical systems theory, we examine how demographic variables (age, gender, professional role) and cultural orientations (language, values, and institutional exposure) mediate perceptions of trust, ethical acceptability, and reliance on AI. Results reveal two key findings:First, cultural factors, particularly language and age, significantly affect AI adoption and perceptions of reliability with older participants reporting higher engagement with AI for educational purposes. Second, ethical judgment about AI use varied across domains, with professional contexts normalizing its role as a pragmatic collaborator while academic settings emphasized risks of plagiarism. These findings extend prior research on culture and technology adoption by demonstrating that AI use is neither universal nor neutral but culturally contingent, domain-specific, and ethically situated. The study highlights implications for AI use in education, professional practice, and global technology policy, pointing at actions that enable usage of AI in a way that is both culturally adaptive and ethically robust.

# Introduction

The swift proliferation of artificial intelligence (AI) technologies in both everyday life and professional environments has raised critical questions regarding their influence on human cognition, behavior, and culture. Much of the existing works have focused on technical innovation, algorithmic performance, and ethical risks (Floridi & Cowls, 2022; Dignum, 2019). However, comparatively less empirical work has examined how AI is adopted across cultures and, more importantly, how this adoption reshapes daily human life strategies. Cross-national evidence demonstrates that perceptions of AI are culturally mediated. For instance, Brauner et al. (2024) found that while German respondents emphasized caution and risk in their assessments of AI, Chinese participants were comparatively more optimistic, highlighting expected societal benefits. Similarly, Yam, Tan, Jackson, and colleagues (2023) argue that cultural traditions, religious orientations, and historical exposure to machines predict differing levels of machine appreciation or aversion between Western and East Asian contexts (Jackson et al., 2023; Lee & Šabanović, 2014; Yam et al., 2023).

This cultural variability extends to organizational and service environments. In the hospitality sector, research shows that trust in AI-driven service robots significantly predicts acceptance, but cultural dimensions such as uncertainty avoidance, power distance, and long-term orientation moderate this relationship (Chi et al., 2023). Likewise, in healthcare, clinicians' adoption of AI is hampered when uncertainty avoidance is high, even in contexts where collectivist values or long-term orientations might otherwise support technological uptake (Krishnamoorthy et al., 2022). In higher education, recent studies demonstrate that generative AI tools can enhance collaboration and creativity, yet their integration also raises concerns about cognitive offloading, which are likely to be interpreted differently across cultural contexts that vary in individualistic versus collectivist orientations (Yusuf, et al. 2024; Iqbal et al., 2025; Gerlich, 2025). Yusuf, Pervin, & Román-González, (2024) examine the usage, benefits, and concerns of generative AI in higher education from a multicultural standpoint. The study found that while generative AI tools can enhance collaboration and creativity, their integration also raises concerns about cognitive offloading, which are likely interpreted differently across cultural contexts that vary in individualistic versus collectivist orientations. Iqbal et al. (2025) investigate the connection between generative AI tool usage and academic achievement in sustainable education, assessing the mediating role of shared metacognition and cognitive offloading among preservice teachers. The study highlights the cognitive implications of AI tool usage and provides actionable recommendations for mitigating potential negative impacts on critical thinking. Gerlich (2025) explores the relationship between AI tool usage and critical thinking skills, focusing on cognitive offloading as a mediating factor. The study investigates how AI tools influence cognitive processes and the extent to which they encourage cognitive offloading, with implications for critical thinking across different cultural contexts

Against this backdrop, the present article seeks to address a persistent gap in literature by presenting findings from a multinational pilot study examining cultural variations in AI use and their implications for human daily life practices. The study aims to understand how cultural dimensions, including language, values, and professional status, mediate the human AI relationship, not only in terms of adoption but also in the ways trust, reliance, and collaboration with AI shape our routine use of AI. Building on these gaps, the present study addresses the following key research question: (1) How do cultural dimensions such as language, age, and institutional context influence the adoption and trust of AI tools in multinational settings?

## Theoretical Framework

This study is grounded in three complementary perspectives: a) cross-cultural psychology, b) digital ethics, and c) sociotechnical systems theory. Drawing on Hofstede's (2010) seminal framework and recent studies (Alsaleh, 2024; Tussey, 2023), culture is conceptualized as a system of shared values, norms, and practices that guide individuals' interpretations and interactions with technology. Alsaleh (2024) defines culture as a dynamic system encompassing shared meanings, practices, beliefs, and values that influence human behavior and are transmitted across generations. Alsaleh's study emphasizes that culture includes both tangible elements like artifacts and technology, as well as intangible components such as beliefs and traditions. From this perspective, cultural factors are expected to influence not only whether AI is adopted but also how it is trusted, evaluated, and incorporated into human activities. For example, previous studies suggest that individuals in high uncertainty avoidance cultures are more likely to prefer predictable, rule-based AI systems over probabilistic or opaque models (Chi et al., 2023). Conversely, in collectivist cultures, social influence and group norms may play a stronger role in shaping AI use, with AI framed as a collaborative or community resource rather than a personal tool (Yam et al., 2023).

The digital ethics literature further highlights that adoption is not a neutral process but one embedded in value laden choices and social contexts (Floridi & Cowls, 2022; Mittelstadt, 2019). Trust emerges as a central mediating factor: cultural values such as power distance, long-term orientation, and collectivism shape the degree of trust invested in AI, which in turn affects the extent to which individuals delegate tasks, rely on AI recommendations, or treat AI as a partner in our daily life activities. For instance, empirical studies of AI in customer service demonstrate that trust is the strongest predictor of adoption, yet cultural dimensions significantly moderate the strength and direction of this effect (Chi et al., 2023).

Finally, sociotechnical systems theory (Bijker, Hughes, & Pinch, 1994) emphasizes the co-evolution of technology and society, suggesting that AI adoption should be theorized as a dynamic, bidirectional process. Individuals adapt AI tools to fit cultural expectations (for instance, by modifying language, interaction styles, or transparency cues), while simultaneously reshaping cultural norms through new practices of AI use. Research on public perceptions of AI across diverse national contexts illustrates this dynamic. Initial cultural framings of AI as assistant, authority, or collaborator strongly shape subsequent usage patterns, yet these frames themselves evolve as users gain familiarity (Chi et al., 2023). The framework therefore anticipates both cross-sectional variation in AI adoption and longitudinal changes in cultural norms as exposure and reliance increase across professional domains.

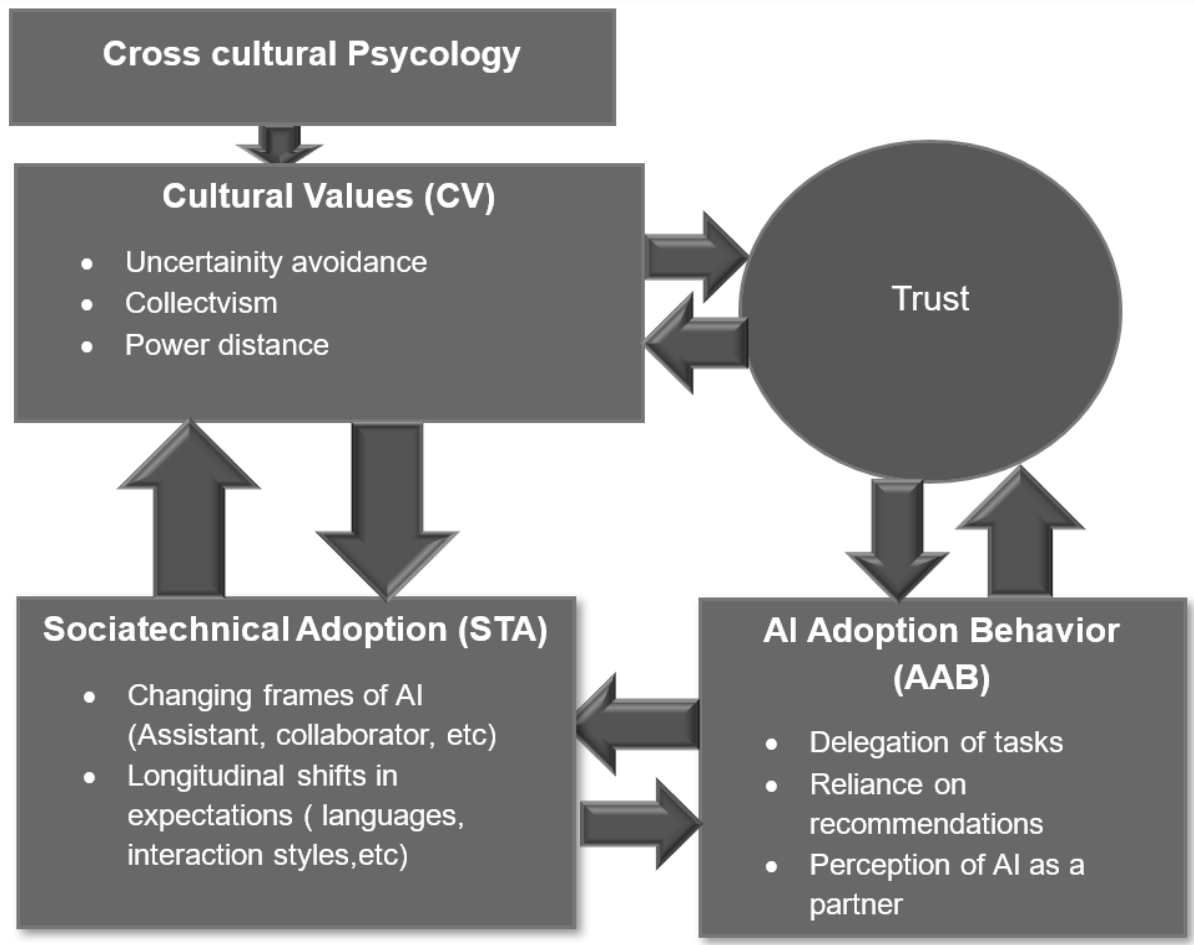

Figure 1: Conceptualization of theoretical framework
Source: Authors own illustration

As shown in Figure 1, the framework illustrates how cultural values, such as uncertainty avoidance, collectivism, and power distance, shape trust in AI, which in turn mediates both socio-technical adoption and adoption behaviors. On the sociotechnical side, cultural orientations influence how AI is framed (e.g., assistant, collaborator, or partner) and how expectations evolve over time regarding language and interaction styles. On the behavioral side, trust conditions practice such as delegating tasks, relying on AI recommendations, and perceiving AI as a partner. Together, the model highlights that AI adoption is not a purely technical process but one deeply embedded in cultural psychology, where trust serves as the bridge linking values, adaptation, and behavior in multinational contexts.

## Materials and Methods

### Research Design

This study is part of a longitudinal pilot study where three waves were conducted over a period of five months (Author, 2025, anonymized) and it employed largely a quantitative pilot design, integrating quantitative survey measures and behavioral observations to capture both broad patterns and in-depth insights. Such approaches are widely recommended in cross-cultural technology adoption research because they allow for triangulation across multiple data sources, thereby strengthening validity and depth (Creswell & Clark, 2017; Venkatesh et al., 2013). Data were drawn from one key primary source. Raw survey responses collected as part of the I-ETOS pilot study on AI and culture, which covered 21 participants representing 9 nationalities across multiple continents. By combining Likert-scale survey items with qualitative self-assessments and observational notes, the study sought to capture not only *whether* and *to what extent* AI tools were adopted, but also *how* cultural dimensions shaped their use in daily contexts.

The study encompasses people from mainly nine culturally distinct national settings (1. Albanian from Kosovo; 2. Belarus; 3 Belgian; 4. Brasilian; 5. German; 6. Egyptian; 7. Nigerian; 8. South African; 9. Thai) (see Figure 2). This purposive selection was informed by prior scholarship indicating that variation along Hofstede's cultural dimensions (e.g., individualism–collectivism, uncertainty avoidance, power distance) significantly shapes technology perceptions and adoption behaviors (Hofstede et al., 2010; House et al., 2004; Taras et al., 2010). By sampling across both Western and non-Western contexts, the study was positioned to generate comparative insights into how culture mediates AI related implementation.

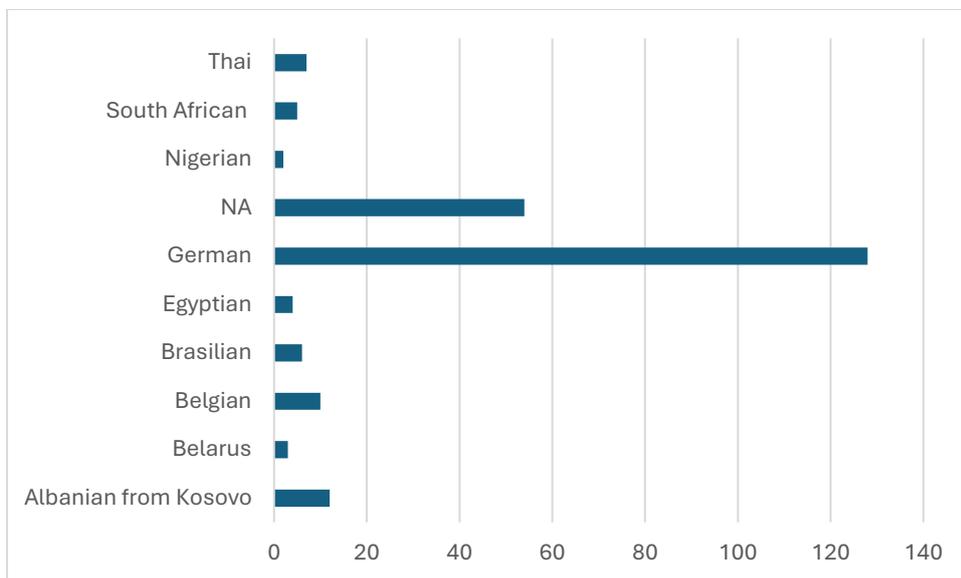

**Figure 2**: Nationality distribution in Study 1.
**Source**: Authors own illustration based on collected data

The overall sample in the study wave 1 spanned mainly four continents (Europe, Africa, Asia, and South America) highlighting the multicultural composition of the study group (see Figure 3)

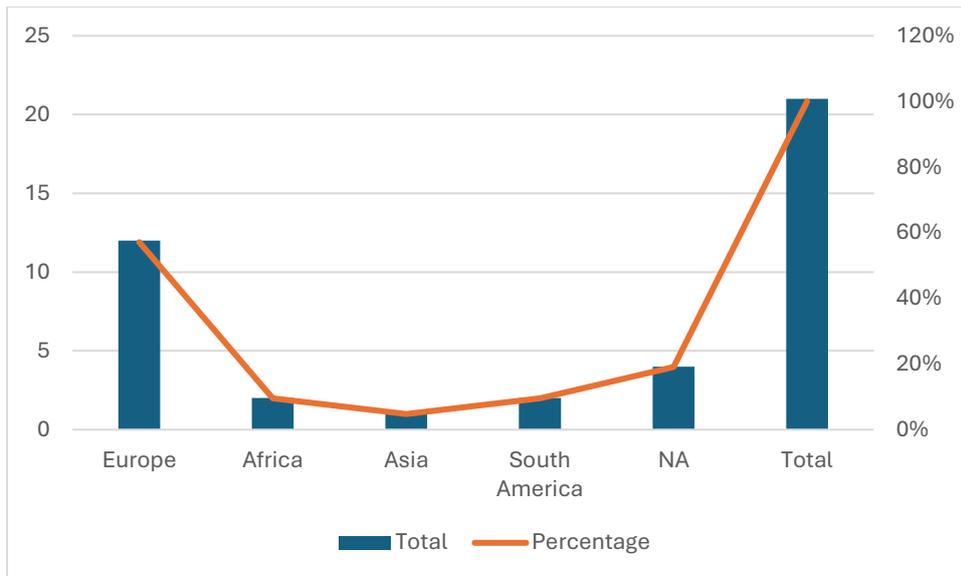

**Figure 3**: Regional participants distribution in Study 1.
**Source**: Authors own illustration based on collected data

### Participants and Proceeding

The survey itself takes approximately 20 minutes to complete, and we have obtained informed consent from all participants before starting. A total of 21 participants participated in the first wave, and they possessed a combination of prior academic institutions, professional associations, and digital networks. This sample emphasized diversity across gender, age, occupation, and educational background (level of education) to reflect heterogeneous perspectives. 42.9% of participants are youthaged below 30 years old ) whiles the rest are adult who are between 30-69 years old and we had no participant with 70 or 70+ years old,52.4% are men vs 47.6 women, 57.2 are not students ( teaching & administration) while the rest are students , 57.1% possess a postgraduate education level ( Masters or equivalent , PhD and Habilitation) while 32.3% possess non post graduate diploma ( High school diploma or Bachelor degrees ) and 4.8 % are classified under Dipl. Ing. (FH). Such demographic and occupational diversity was considered essential, as prior research demonstrates that age, professional identity, and prior digital experience can significantly moderate technology adoption (Venkatesh et al., 2012; Marangunić & Granić, 2015; Cai et al., 2017). Comparative analysis between different participants demographics further allowed exploration of institutional culture as a potential moderating factor in AI adoption. For instance, institutional affiliation or occupation has been shown to shape not only access to technological resources but also normative expectations regarding responsible use (Wang & Siau, 2018; Raisch & Krakowski, 2021).

**Measures**

To ensure comparability and construct validity, multiple measures were employed:

- **Cultural Values.** Cultural orientations were assessed using Hofstede's *Values Survey Module* (VSM 2013). This instrument, validated across numerous multinational studies, captures key cultural dimensions such as power distance, individualism–collectivism, masculinity–femininity, uncertainty avoidance, and long-term orientation (Hofstede et al., 2010; Taras et al., 2010).

- **AI Adoption.** Adoption was measured through self-reported frequency of AI tool usage, perceived usefulness, and trust. These constructions were operationalized following the Unified Theory of Acceptance and Use of Technology (UTAUT2) framework, which emphasizes performance expectancy, effort expectancy, social influence, and trust as predictors of adoption (Venkatesh et al., 2012; Dwivedi et al., 2023).

**Data Analysis**

The data were analyzed using Jamovi 2.6.44, a widely recognized open-source statistical platform for social science research (The Jamovi Project, 2023). Descriptive statistics, including frequencies, means, standard deviations, and percentages, were employed to provide a comprehensive overview of participants' demographic profiles, educational backgrounds, and professional experiences. Such descriptive analyses are essential in exploratory and cross-cultural studies, as they facilitate a nuanced understanding of sample diversity and contextual variation (Field, 2024; Van de Vijver, & Leung, 2021).

In addition, cross-tabulations were conducted to examine associations between categorical variables and to explore the influence of participants' professional status on their approaches to AI use (see Table 5). Cross-tabulation is particularly valuable in exploratory studies with smaller, heterogeneous samples, as it enables the identification of patterns and relationships that may inform theoretical and practical insights (Agresti & Kateri, 2025; Babbie, 2013). Moreover, a Chi-square test was performed to establish relationship between participants and AI use for study and educational purposes. Together, these analytic strategies provided a robust foundation for examining the role of cultural dimensions, trust, and sociotechnical factors in shaping AI adoption across diverse multinational contexts.

## Results

### Participant Demographics

The study sample (n = 21) was characterized by substantial cultural and demographic diversity, encompassing nine nationalities across four continents (Europe, Africa, Asia, and South America), see Table 1. The largest subgroup comprised German participants (38.1%), followed by individuals from Nigeria, South Africa, Thailand, Brazil, and Egypt. The gender distribution was nearly balanced (52.4% men; 47.6% women). In terms of age, 42.9% were youth (below 30 years), while the remainder were adults aged 30–69 years; no participants aged 70 years or above were included. Educational attainment was notably high: 57.1% held postgraduate qualifications (master's degree or equivalent, PhD, or Habilitation), 32.3% held undergraduate or high school diplomas, and 4.8% were classified under Dipl. Ing. (FH). Regarding professional status, 57.2% were non-students (primarily in teaching and administration), with the remainder comprising active students.

Table 1. Participants demographics

| Characteristic | Category | Frequency (f) | Percentages % |
|---|---|---|---|
| **Age** | Youth (<30 years) | 9 | 42.9 |
| | Adults (30–69 years) | 12 | 57.1 |
| | Older adults (≥70 years) | 0 | 0.0 |

| Characteristic | Category | Frequency (f) | Percentages % |
|---|---|---|---|
| **Gender** | Men | 11 | 52.4 |
| | Women | 10 | 47.6 |
| **Nationality** | German | 8 | 38.1 |
| | Nigerian, South African, Thai, Brazilian, Egyptian | 5 | 23.8 |
| | Other (Kosovar Albanian, Belarusian, Belgian) | 8 | 38.1 |
| **Occupation** | Non-students (Teaching + Admin) | 12 | 57.2 |
| | Students | 9 | 42.8 |
| **Education Level** | Non-postgraduate (High school + Bachelor) | 7 | 33.3 |
| | Postgraduate (Master's + PhD + Habilitation) | 13 | 61.9 |
| | Other (Dipl. Ing. FH) | 1 | 4.8 |

**Note.** Percentages are based on the total sample size (n = 21). Nationalities are grouped for clarity: the largest subgroup was German participants, followed by a mix of African, Asian, and South American participants, and a smaller group from other European countries.

**Source**: Authors own illustration based on collected data

### AI Knowledge and Usage Patterns

Following the demographics we obtained a general understanding about the background in AI Knowledge and its Usage, see Table 2. Most participants (81%) reported having prior knowledge of AI. In terms of tool adoption, ChatGPT was the most widely used platform (85.7%), followed by Microsoft Copilot (38.1%), Google Gemini (23.8%), and Claude (19%). A range of other tools, including AI applications in Adobe Creative Cloud, DeepL, DeepSeek, Goodnotes, Elicit, Research Rabbit, TextSynth, and Siri, were also reported, collectively accounting for 38.1% of use. AI engagement extended across multiple domains: 76.2% of participants reported frequent use in private contexts, an equal proportion (76.2%) in professional contexts, and 71.4% in educational activities. Patterns of intensity further revealed that a majority (52.4%) engaged with AI tools on a daily basis, underscoring the pervasiveness of AI in participants' everyday practices.

Table 2. AI Knowledge and Usage Patterns

| Characteristic | Category | Frequency (f) | Percentages % |
|---|---|---|---|
| **AI Knowledge** | Prior knowledge in AI | 17 | 81.0 |
| | No prior knowledge | 4 | 19.0 |
| **AI Tools Used** | ChatGPT | 18 | 85.7 |
| | Microsoft Copilot | 8 | 38.1 |
| | Google Gemini | 5 | 23.8 |
| | Claude | 4 | 19.0 |
| | Other tools (Adobe CC AI, DeepL, Deepseek, Goodnotes AI, ML tools, Elicit, Research Rabbit, TextSynth, Siri) | 8 | 38.1 |
| **Contexts of AI Use** | Private | 16 | 76.2 |
| | Professional | 16 | 76.2 |
| | Educational | 15 | 71.4 |
| **Frequency of AI Use** | Several times a day and once a day | 11 | 52.4 |
| | Several times a week and once a week | 8 | 38.1 |
| | Other | 2 | 9.5 |

**Note.** Percentages are based on the total sample size (n = 21). Multiple responses were possible for AI tools. Reported durations indicate most participants had between one and two years of experience; exact distributions are approximate in some categories due to reporting format.

**Source**: Authors own illustration based on collected data

### Perceptions of AI

Perceptions of AI varied considerably among participants, see Table 3. While the majority regarded AI as moderately reliable, notable skepticism remained. Concerns regarding data protection were particularly pronounced, with participants reporting a relatively high level of unease (mean = 5.9 on a scale from 1 = *protected* to 8 = *unprotected/unsafe*). Ethical evaluations of AI also revealed contextual differences: although AI was generally accepted for private and professional use, its application in academic settings, particularly in thesis writing, was viewed as problematic. These findings indicate that trust in AI is not uniform but shaped by concerns over security and ethical appropriateness, which vary by domain of application.

Table 3. Perceptions of AI (n=21)

| Dimension | Findings | % / M (SD) |
|---|---|---|
| **Perceived Reliability** | AI considered moderately reliable (mean = 4.1 on 1–8 scale) | M = 4.1 (±1.5) |
| **Data Safety** | Concerns about data protection; many felt unsafe (mean = 5.9 on 1–8 scale, where 1 = protected, 8 = unsafe) | M = 5.9 (±2.2) |
| **Ethical Use** | Use in private contexts not widely considered problematic | 76.2% said "not cheating" |
| | Use in professional contexts not widely considered problematic | 57.1% said "not cheating" |
| | Use in academic theses often seen as problematic (plagiarism risk) | 66.7% said "cheating" |

**Note.** Reliability and safety were measured on 8-point Likert scales (1 = low, 8 = high). Percentages reflect categorical survey responses. While AI was broadly accepted for private and professional tasks, strong concerns remained about its ethical use in academic work.

**Source**: Authors own illustration based on collected data

### Cultural Dimensions of AI Use

Cultural Dimensions of AI Use were strongly acknowledged by participants, with 81% agreeing that culture significantly shapes patterns of AI engagement. Language emerged as a particularly salient factor (71.4%), with participants emphasizing the role of academic jargon, field-specific terminology, and distinctions between native and non-native language use in shaping the effectiveness of AI outputs. Age was almost universally recognized as influential (95.2%), highlighting generational differences in adoption practices. By contrast, gender and religion were perceived as having comparatively limited impact. Importantly, the majority of participants (78.9%) reported that AI frequently produces responses that reflect culture-specific orientations, underscoring the contextual embeddedness of AI interaction

Table 4. Cultural Dimensions of AI Use

| Dimension | Findings / Categories | Frequency (f) | Parcentages (%) |
|---|---|---|---|
| **Cultural Influence** | Agree that culture influences AI use | 17 | 81.0 |
| | Disagree | 4 | 19.0 |
| **Cultural Sensitivity** | Agree that AI depicts responses that are culture specific | 16 | 78.9 |
| | Disagree | 5 | 21.1 |
| **Language Influence** | Agree that language influences AI use | 15 | 71.4 |
| **Language Aspects** | Disagree | 6 | 28.6 |
| | Native vs. non-native usage | 11 | 52.4 |
| | Field-specific jargon | 14 | 66.7 |
| | Academic vs. non-academic language | 14 | 66.7 |
| | Other (tone, mannerisms, style) | 2 | 9.5 |
| **Age Influence** | Agree | 20 | 95.2 |
| | Disagree | 1 | 4.8 |
| **Gender Influence** | Agree | 10 | 47.6 |
| | Disagree | 11 | 52.4 |
| **Religious Influence** | Agree | 5 | 25.0 |
| | Disagree | 15 | 75.0 |

**Note.** Percentages are based on total participants (n = 21), except for religion where n = 20. Multiple responses were possible for language aspects. While most participants acknowledged cultural and linguistic influence on AI use, many reported AI struggles to capture cultural nuance accurately

**Source**: Authors own illustration based on collected data

**Crosstabulation: Age vs AI use for studies and educational purposes**

Most respondents (15 out of 21) belonged to age group 1, while the remaining six participants were in age group 2. The distribution of responses indicates that younger individuals (age group 1) were more likely to fall within the lower AI use categories (1–3), whereas older individuals (age group 2) were more represented in the higher AI use categories (4–5). This pattern suggests a tendency for older respondents to report greater use of AI for educational purposes compared to their younger counterparts.

Table 5: Crosstabulation: Age vs AI use for studies and educational purposes

| What is your age? | Do you use AI for studies and educational purposes? | | Total |
|---|---|---|---|
| | 1 | 2 | |
| G1 | 1 | 0 | 1 |
| G2 | 8 | 0 | 8 |
| G3 | 3 | 1 | 4 |
| G4 | 3 | 2 | 5 |
| G5 | 0 | 3 | 3 |
| Total | 15 | 6 | 21 |

Source: Authors owns illustration based on collected data

According to table 5 above, A cross-tabulation between age and AI use for educational purposes suggested potential differences in usage across age groups. Most younger respondents reported lower levels of AI use, while older participants tended to report higher levels. A chi-square test of independence has been performed to determine whether this pattern is statistically significant (see Table 6). According to table 6, $p = 0.022$ was obtained, which is less than 0.05, which means the result is statistically significant at the 5% level. This allows us to reject the null hypothesis that there is no significant relation between age and AI use for educational purposes.

Table 6. Chi-Square Test Results

| Statistic | Value | Interpretation |
| --- | --- | --- |
| χ² | 11.4 | Measures the difference between observed and expected frequencies. |
| df | 4 | Degrees of freedom: (rows – 1) × (columns – 1) = (5 – 1) × (2 – 1) = 4 |
| p-value | 0.022 | Probability that the observed pattern happened by chance. |
| N | 21 | Sample size |

Source: Authors' owns illustration based on the collected data

From the table 6 above, it is obvious that there is a statistically significant association between age and AI use for studies, χ² (4, N = 21) = 11.4, p = 0.022. In order words, older participants tend to report higher levels of AI use for educational purposes, while younger participants tend to report lower levels.

## Discussion

The findings of this pilot study underscore the salience of cultural dimensions in shaping AI adoption and practices, aligning with prior scholarship emphasizing that technological uptake is never culturally neutral (Hofstede et al., 2010; Dignum, 2019; Floridi & Cowls, 2022). The results provide nuanced insights into how demographic diversity, knowledge and usage patterns, perceptions of trust and ethics, and cultural framings interact to influence the ways humans engage with AI for cognitive and practical tasks.

### Participant Diversity and Cultural Exposure

The sample, spanning nine nationalities across four continents, reflects the multicultural realities of AI adoption. Although small, this diversity allowed us to observe differences in how cultural background shapes AI engagement. Previous studies have shown that exposure to multicultural environments fosters greater sensitivity toward technological diversity and ethical considerations (Taras et al., 2010; Stahl et al., 2017). In line with Yam et al. (2023), our findings suggest that younger participants and those with postgraduate training were more open to adopting AI, a trend consistent with generational shifts toward digital fluency and technological optimism.

### AI Knowledge and Usage Patterns

The widespread use of ChatGPT and similar generative AI tools among participants mirror global adoption patterns, where large language models have become the dominant entry point for AI interaction (Achiam et al., 2023; Dwivedi et al., 2023). According to scholars such as Shneiderman (2020; 2022), the accessibility of natural language interfaces reduces barriers to entry and democratizes problem-solving capacities. However, the simultaneous use of specialized tools (e.g., Adobe AI, DeepL, Research Rabbit) indicates that AI adoption is contextual and task-specific, supporting findings by Akhlaghpour et al., (2025) that functional alignment with user needs strongly predicts adoption.

### Age and Patterns of AI Use

The quantitative results revealed a statistically significant association between age and AI use for educational purposes (χ² (4, N = 21) = 11.4, p = .022), indicating that older participants were more likely to report higher engagement with AI tools for study-related activities. This finding aligns with previous research suggesting that age can shape both the motivation and confidence with which individuals adopt emerging technologies (Venkatesh et al., 2012; Hargittai, 2020). One possible interpretation is that older respondents, who often occupy professional or postgraduate academic roles, perceive AI as a pragmatic aid that enhances productivity and supports lifelong learning. Conversely, younger participants, typically students still navigating institutional expectations, may exhibit more cautious or skeptical attitudes toward AI, especially within educational settings where its use is closely tied to debates about plagiarism and academic integrity.

This age-related variation in adoption resonates with cross-cultural theories emphasizing that demographic and cultural orientations jointly mediate technological perception (Hofstede et al., 2010; Taras et al., 2010). In this sense, age functions not merely as a demographic variable but as a proxy for cultural experience, institutional exposure, and ethical framing of AI use.

### Perceptions of AI: Trust, Risk, and Ethics

While most participants viewed AI as moderately reliable, persistent skepticism, particularly regarding data protection, highlights an enduring trust gap. Previous research has shown that trust in AI is mediated by both cultural orientation and institutional trust in data governance (Mittelstadt, 2019; Chi et al., 2023). In our sample, participants distinguished between acceptable use in professional or private domains and more contentious use in academic contexts, particularly thesis writing. This reflects what Stahl et al. (2021) describe as situational ethics, where context determines whether AI use is perceived as augmentation or misconduct. The pragmatic acceptance of AI in workplaces, contrasted with strict concerns about plagiarism in education, supports earlier findings that ethical boundaries are socially and institutionally constructed (Jobin et al., 2019).

### Cultural Dimensions of AI Use

A strong majority (81%) acknowledged that culture influences AI adoption, particularly through language and age-related differences. Language, including native vs. non-native fluency, was seen as a major factor in AI's effectiveness, a finding consistent with Spencer-Oatey (2008) on the role of communication norms in shaping intercultural technology use. Recent studies confirm that large language models often reproduce linguistic biases, amplifying barriers for non-native users (Weidinger et al., 2021; Hovy & Prabhumoye, 2021). Age differences were also widely recognized, echoing prior evidence that digital natives display greater confidence in experimenting with AI compared to older cohorts who may adopt more cautiously (Vaportzis et al., 2017; Ransbotham et al., 2020). Interestingly, gender and religion were perceived as less influential, contrasting with some earlier work that emphasized their role in shaping digital adoption patterns (Cai et al., 2017). This may reflect the institutional academic context of our sample, where professional identity outweighs other cultural markers.

### Integrating Cultural Context into AI Adoption

Taken together, these findings reinforce the core argument of this study. AI adoption and its impact on our daily life practices are deeply shaped by cultural dimensions. According to cross-cultural psychology, cultural values such as collectivism, uncertainty avoidance, and power distance inform not only attitudes toward technology but also the strategies individuals employ when solving problems (Hofstede et al., 2010; Taras et al., 2010; Yam et al., 2023). In our study, participants explicitly recognized language and age as shaping factors, while institutional exposure mediated openness to cultural variability. This suggests that effective AI integration requires attention not only to technical performance but also to cultural adaptability and ethical guidance.

**Ethical Considerations**

The integration of AI into our routine practices inevitably raises complex ethical questions. Findings from the pilot study illustrate this tension. While 57.1% of respondents rejected the notion that the use of AI in professional tasks constitutes "cheating," a majority (66.7%) considered AI-assisted thesis writing to be a form of plagiarism if not explicitly cited. These results suggest that ethical frameworks surrounding AI are highly contextual, with professional pragmatism often prioritized in workplace settings, whereas academic environments maintain stricter standards of integrity.

This context dependence aligns with broader debates in the digital ethics literature, where scholars such as Floridi and Cowls (2022) argue that ethical evaluations of AI cannot be reduced to universal principles but must instead account for domain-specific norms and cultural values. For example, research on generative AI in higher education shows that students and faculty differ in their ethical assessments depending on whether AI is framed as a learning aid, a writing tool, or a substitute for individual effort (Haroud, & Saqri, 2025; Johnston et al.,2024; Kasneci et al., 2023). Several studies show that students and faculty differ in their ethical assessments of generative AI depending on the tool's framing: students are more accepting when AI is presented as a learning aid (e.g., for grammar, summaries) but are much less comfortable when it is used as a writing tool or to substitute substantial portions of individual effort (Haroud, & Saqri, 2025; Johnston et al., 2024). For example , the study by Johnston et al., (2024) , which is a large-scale survey (n ~2555 students) indicates that students differentiate ethical acceptability depending on how the AI is used where many are okay with tools like Grammarly or using generative AI for grammar help / small edits, but much less comfortable when it comes to using it to produce whole essays or doing work that substitutes the student's own effort. On the other hand, research by Haroud, & Saqri, (2025) found significant differences between students and teachers in how they perceive generative AI. Students tend to view these tools more positively (i.e. for support, creativity, ease), whereas teachers are more cautious, especially regarding replacing parts of teaching, risks to skills like critical thinking, and ethical implications. Similarly, studies in organizational contexts demonstrate that employees are more likely to view AI as an efficiency-enhancing collaborator rather than as a threat to professional integrity, particularly in cultures that emphasize long-term orientation and pragmatism (Raisch & Krakowski, 2021).

Cross-cultural perspectives further complicate these debates. Yam et al. (2023) show that cultural traditions influence ethical judgments about algorithmic and robotic systems. In collectivist societies, the legitimacy of AI use is more closely tied to social consensus, whereas in individualist cultures, personal responsibility and transparency dominate ethical reasoning (Barnes et al., 2024; Cao, & Meng, 2025; Dang, & Li 2025). For instance, the study by Barnes, Zhang, & Valenzuela (2024) suggests how cultural identity influences perceptions of AI (e.g. people from collectivist cultures may see AI more as part of a social system, less as an infringement on personal autonomy, and more in terms of its alignment with collective or normative expectations). Individualists, by contrast, are more concerned about autonomy, personal rights, privacy, and transparency. This helps explain why the same AI tool might be considered acceptable in a corporate context in one culture but unethical in an academic setting in another. Moreover, uncertainty avoidance has been linked to stricter ethical stances regarding AI use, with high-uncertainty-avoidance cultures requiring more explicit regulation and disclosure (Chi et al., 2023).

The study's findings also underscore the dynamic nature of ethical frameworks. perceptions of fairness, transparency, and integrity evolve as individuals gain familiarity with AI. What is initially regarded as inappropriate or risky may become normalized over time, echoing sociotechnical systems theory, which emphasizes the co-construction of technological practice and normative expectations (Bijker, Hughes, & Pinch, 1994). However, as Mittelstadt (2019) cautions, normalization without critical oversight risks ethical complacency, highlighting the need for continuous dialogue between policymakers, institutions, and users.

The ethical implications of AI adoption are best understood as situated, culturally mediated, and evolving. Professional pragmatism and academic integrity represent two ends of a spectrum in which AI's legitimacy is negotiated differently across domains, cultures, and institutional settings. Future governance frameworks must therefore remain sensitive to context, embedding ethical guidelines that are flexible enough to adapt to diverse cultural values while robust enough to uphold principles of transparency, accountability, and fairness.

## Conclusion

This pilot study contributes to the growing literature on the intersection of culture and artificial intelligence. By examining nine culturally diverse national contexts, we provide empirical evidence that cultural dimensions, particularly language and age, play a decisive role in shaping both the adoption of AI tools and the ways in which individuals engage with them cognitively. As previous studies have shown (Hofstede et al., 2010; Yam et al., 2023; Taras et al., 2010), cultural orientations strongly mediate technological perceptions, and our findings confirm this dynamic in the realm of AI.

As the chi-square analysis revealed, age significantly influenced AI use in educational contexts, underlining the demographic dimension of cultural variability. Together with language-related differences, this underscores how demographic and cultural dimensions jointly influence the cognitive and ethical framing of AI use.

The ethical dimension further underscores that AI use cannot be disentangled from cultural and institutional contexts. Whereas professional settings often normalize AI as a pragmatic collaborator, academic contexts continue to frame certain uses as plagiarism. According to Floridi and Cowls (2022), ethical frameworks for AI must remain adaptive and domain specific. Our results resonate with earlier evidence from Jobin et al. (2019) and Kasneci et al. (2023) that perceptions of fairness, transparency, and legitimacy evolve with exposure and cultural framing.

Taken together, the study advances three key insights: First, cultural variability is a decisive factor in AI adoption, requiring sensitivity in design and policy. Second, AI's role in our daily practices is inherently bound, highlighting the importance of augmentation rather than replacement. Third, ethical evaluations of AI remain dynamic and culturally mediated, demanding flexible yet robust governance frameworks. As scholars such as Mittelstadt (2019) caution, normalization without critical oversight risks ethical complacency. While this pilot study is limited in scope, it highlights the urgent need for cross-cultural, longitudinal, and mixed-methods research to deepen our understanding of how AI adoption intersects with human cognition and cultural values. Situating AI within the lived realities of diverse cultural contexts, future research can ensure that technological integration advances both efficiency and human flourishing.

## Limitations and Future Perspectives

Like all pilot studies, this research is subject to several limitations that warrant caution in interpretation. First, the sample size (n = 21 in the WodETOS pilot data) and its concentration within a single academic institution (ECRI) constrain the generalizability of findings. While pilot studies are valuable for generating exploratory insights, small samples limit statistical power and increase the risk of overestimating effects (Julious, 2005; van Teijlingen & Hundley, 2002). Expanding future research to include larger, more demographically and institutionally diverse cohorts will be essential to confirm and extend the observed patterns. However, the significant association suggests meaningful patterns that warrant further cross-cultural and longitudinal investigation.

Second, the cross-sectional design of the study captures only a snapshot of participants' attitudes and behaviors. Given that perceptions of AI evolve with familiarity and societal exposure (Raisch & Krakowski, 2021; Long & Magerko, 2020), future research should employ longitudinal tracking to investigate how cultural factors and daily work routine strategies shift over time. Such an approach would help capture the dynamic nature of AI adoption and the co-evolution of cultural norms with technological practice, as theorized in sociotechnical systems frameworks (Bijker, Hughes, & Pinch, 1994).

Third, while the study combined survey data with limited qualitative self-assessments, a richer understanding of cultural nuances requires more extensive use of qualitative methods such as in-depth interviews, focus groups, or ethnographic observation. Previous work in cross-cultural psychology emphasizes that surveys may obscure subtle cultural meanings that can only be elicited through dialogic and interpretive methods (Smith, Fischer, & Vignoles, 2013). Incorporating such qualitative approaches would enable researchers to better capture how values, language, and institutional norms mediate AI adoption in everyday life contexts.

Fourth, the absence of experimental designs restricts the ability to establish causal relationships. While this study identified associations between cultural dimensions and AI trust, future research should incorporate controlled experiments to directly test how AI use influences accuracy, creativity, and cognitive load across cultural groups. Experimental paradigms have proven especially useful for isolating the psychological mechanisms through which culture interacts with technology adoption (Taras et al., 2012; Chua, Morris et al., 2012; Taras et al., 2012. For instance, studies by Chua, Morris, and Mor (2012) as well as Taras et al., (2012) highlight that experimental approaches are particularly effective in uncovering the underlying psychological processes that shape how culture influences the adoption of technology.

Finally, this study focused primarily on academic and professional contexts, leaving other domains underexplored. Future investigations should examine AI adoption in areas such as healthcare, public administration, and everyday decision-making, where cultural values may play a different or more pronounced role. Cross-sectoral comparisons could shed light on whether cultural sensitivities are domain-specific or represent more generalized orientations toward AI. These limitations highlight the need for multi-method, cross-national, and longitudinal research programs that move beyond exploratory pilot work. By combining quantitative and qualitative methods, experimental and observational designs, and culturally diverse samples, future studies will be better positioned to capture the complexity of how AI adoption interacts with cultural dimensions to shape human problem-solving.